\documentclass[lettersize,journal]{IEEEtran}
\usepackage{amsmath,amsfonts}
\usepackage{algorithmic}
\usepackage{algorithm}
\usepackage{array}
\usepackage{courier}
\usepackage{textcomp}
\usepackage{stfloats}
\usepackage{booktabs}
\usepackage{url}
\usepackage{verbatim}
\usepackage{makecell}
\usepackage{graphicx}
\usepackage{caption}
\usepackage{subcaption}
\usepackage{cite}
\usepackage{multirow}
\usepackage{graphicx} 
\usepackage{mathtools}
\usepackage{color, colortbl}
\usepackage[dvipsnames]{xcolor}
\definecolor{lightred}{HTML}{FFCCCB}
\definecolor{lightgreen}{HTML}{D1FFBD}
\providecommand{\gz}[1]{\textcolor{blue}{{#1}}}
\providecommand{\GZ}[1]{\textcolor{blue}{[{\bf #1}]}}
\providecommand{\jd}[1]{\textcolor{brown}{{#1}}}
\providecommand{\JD}[1]{\textcolor{brown}{[{\bf #1}]}}
\providecommand{\zd}[1]{\textcolor{red}{{#1}}}
\providecommand{\ZD}[1]{\textcolor{red}{[{\bf #1}]}}
\begin{document}
\title{Cacophony: An Improved Contrastive \\ Audio-Text Model}

\author{Ge Zhu,~\IEEEmembership{Graduate Student Member,~IEEE,} Jordan Darefsky, Zhiyao Duan,~\IEEEmembership{Member, IEEE}\\ University of Rochester, USA 
\thanks{Ge Zhu and Zhiyao Duan are with the Department of Electrical and Computer Engineering at the University of Rochester. Jordan Darefsky is with the Department of Computer Science at the University of Rochester. This works is partially supported by the New York State Center of Excellence in Data Science and the University of Rochester Goergen Institute for Data Science seed funding program.}
}



\providecommand{\gz}[1]{\textcolor{black}{{#1}}}
\providecommand{\GZ}[1]{\textcolor{black}{[{\bf #1}]}}
\providecommand{\jd}[1]{\textcolor{cyan}{{#1}}}
\providecommand{\JD}[1]{\textcolor{cyan}{[{\bf #1}]}}
\providecommand{\zd}[1]{\textcolor{magenta}{{#1}}}
\providecommand{\ZD}[1]{\textcolor{magenta}{[{\bf #1}]}}
\maketitle

\begin{abstract}
Despite recent advancements, audio-text models still lag behind their image-text counterparts in scale and performance. 
In this paper, we propose to improve both the data scale and the training procedure of audio-text contrastive models. Specifically, we craft a large-scale audio-text dataset containing 13,000 hours of text-labeled audio, using pretrained language models to process noisy text descriptions and automatic captioning to obtain text descriptions for unlabeled audio samples. We first train on audio-only data with a masked autoencoder (MAE) objective, which allows us to benefit from the scalability of unlabeled audio datasets. 
We then train a contrastive model with an auxiliary captioning objective with the audio encoder initialized from the MAE model. 
Our final model, which we name Cacophony, achieves state-of-the-art performance on audio-text retrieval tasks, and exhibits competitive results on the HEAR benchmark and other downstream tasks such as zero-shot classification. 


\end{abstract}


\section{Introduction}

Machine audition~\cite{wang2010machine} involves developing algorithms and systems for machines to analyze and understand sound, covering tasks such as audio tagging, acoustic scene classification, music classification, and sound event detection. In recent years, there has been a general shift away from individual audio pattern recognition tasks and toward general-purpose audio representations pretrained on large-scale audio datasets. 
Pretrained Audio Neural Networks (PANNs) ~\cite{kong2020panns} have played a significant role in this shift by demonstrating their versatility across various tasks and outperforming many advanced systems via fine-tuning. 

Modern approaches aim for robust performance in general-purpose audio understanding tasks without the need for task-specific fine-tuning, which offers more flexibility.
These methods approach general audio understanding by linking text and audio modalities, referred to as \textit{audio-text models}.
One approach to linking text and audio is through generating response text given a combination of an audio prompt and a text prompt~\cite{gardner2023llark,gong2023listen,deshmukh2023pengi}.
For instance, Pengi~\cite{deshmukh2023pengi} converts audio classification, retrieval, captioning, and audio question answering into a text generation task using audio and task-specific text prompts.
Similarly, Qwen-Audio~\cite{chu2023qwen} addresses a variety of audio tasks through text generation but distinguishes itself by using text prompts consisting of hierarchical tag sequences inspired by Whisper~\cite{radford2023robust}.
Both Pengi and Qwen-Audio support multiple closed-ended and open-ended audio tasks without the need for additional fine-tuning or task-specific extensions of the architecture.
Another method of linking text and audio is contrastive learning.
Pretrained contrastive models can also be applied directly to various downstream tasks without fine-tuning.
For instance, contrastive models can be used for retrieval and classification by assigning a score that identifies the most probable text (or class label) from a predefined set of choices for a given audio input.
Moreover, the learned audio-text representations from contrastive models offer the flexibility to use one modality during training and the other at inference, which can be applied in text-to-audio generation~\cite{huang2023make,liu2023audioldm} and language-guided source separation~\cite{liu2023separate,dong2022clipsep}.



\begin{figure}[!t]
\centering
\includegraphics[width=0.9\columnwidth]{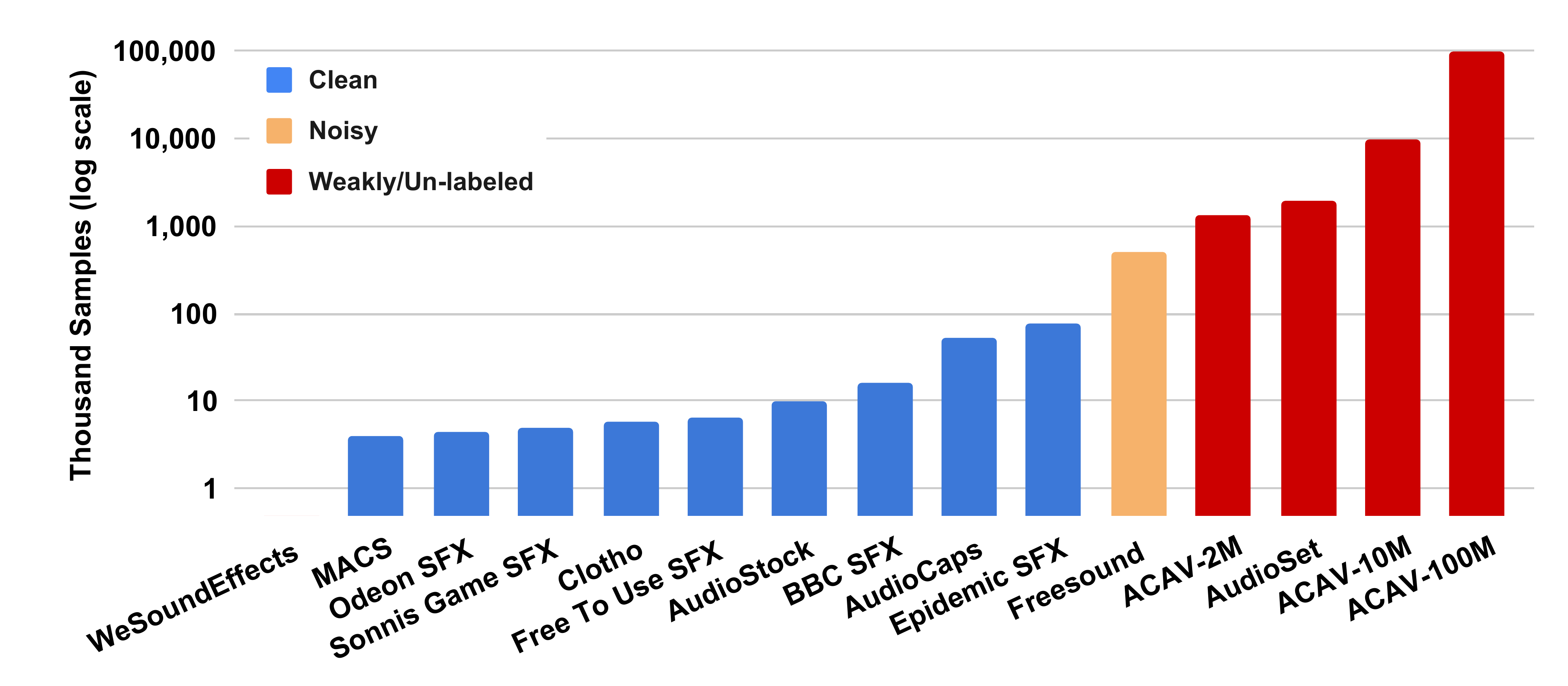}
\caption{A bar graph of the number of samples from commonly-used public audio datasets.}
\label{fig:audiodata}
\end{figure}

In this paper, we focus on improving contrastive audio-text models for general sounds by addressing two critical limitations in existing research: insufficient dataset scale and the vanilla contrastive training techniques.
To tackle the dataset scale issue, we first examine the publicly available audio datasets. 
These can be categorized into three types based on label granularity: clean-labeled, noisy-labeled, and weakly-labeled or unlabeled, as illustrated in Fig.~\ref{fig:audiodata}. 
Clean-labeled datasets, while high-quality, are limited in size. Noisy-labeled datasets offer more samples but include extraneous details. 
The largest category by far is weakly-labeled or unlabeled data, which provides little to no textual information.
This scarcity of high-quality labeled data presents a significant challenge in audio-text contrastive learning, especially when compared to recent advancements in image-text models. 
For context, while the largest public audio-text datasets contain less than 100,000 pairs~\cite{kim2019audiocaps,drossos2020clotho}, image-text models like CLIP~\cite{radford2021learning} and SigLip~\cite{zhai2023sigmoid} utilize 400 million and 3.6 billion pairs respectively.

Previous works have attempted to address this data scarcity issue by collecting data from various sources and applying natural language processing techniques to clean or filter noisy captions.
For instance, Huang et al.~\cite{huang2022mulan} collected approximately 44 million 30-second music clips, applying a pretrained classifier and rule-based filtering to clean associated metadata. 
This process intensively reduced the dataset to 2 million music-text pairs. 
Notably, they found that models trained on large scale unfiltered audio-text data performed comparably to those trained on filtered data in music tagging and retrieval tasks, suggesting that data quantity might be as crucial as quality in this domain.
Wu et al.~\cite{laionclap2023}, in collaboration with Large-scale Artificial Intelligence Open Network (LAION), curate the LAION-Audio dataset with 630K audio-text pairs together with 2 million Audioset~\cite{gemmeke2017audio} clips recaptioned with keyword-to-caption (K2C) augmentation to train LAION-CLAP.
The K2C method uses a pretrained language model to generate captions from tags.
However, the captions produced via K2C are restricted to the objects defined by the tags, offering limited descriptive details.
K2C also risks making incorrect assumptions or introducing biases highlighted in~\cite{laionclap2023}.
Mei et al.~\cite{mei2023wavcaps} propose a multi-stage data filtering pipeline and utilize ChatGPT for cleaning text descriptions. 
Their contrastive language-audio pretraining (CLAP) models, trained on their WavCaps dataset, demonstrate superior performance in audio-text retrieval tasks compared to LAION-CLAP, despite a smaller-scale dataset.
Such a text filtering pipeline reduces the amount of audio data, which could potentially result in reduced generalization.

In addition to dataset scale, there is a need for novel neural architectures and training strategies tailored to model audio structures more effectively. 
For instance, LAION-CLAP~\cite{laionclap2023} investigates different choices of audio/text encoders, demonstrating superior performance with the hierarchical token semantic audio transformer (HTSAT)\cite{chen2022hts} for audio encoding and the Robustly optimized BERT\footnote{acronym for `Bidirectional Encoder Representations from Transformers'} approach (RoBERTa)\cite{liu2019roberta} for text encoding.
In addition, LAION-CLAP proposes feature fusion for audio inputs with variable-length.
In concurrent work, fast language-audio pretraining (FLAP)~\cite{yeh2023flap}, inspired by fast language-image pretraining (FLIP)~\cite{li2023scaling}, proposes masking and removing a significant portion of spectrogram patches. 
FLAP also incorporates a reconstruction loss during the contrastive training on these masked spectrogram patches, although the improvements are modest. 

In this paper, we investigate several strategies to improve the audio-text models, informed by the aforementioned challenges. 
For dataset creation, we collect a large-scale audio-text dataset and expand and refine its text descriptions.
For audio recordings with weak or no labels, we utilize an automatic audio captioning model to obtain synthetic captions. 
For audio paired with noisy descriptions, we use large language models (LLMs) to generate several cleaned captions for each audio clip. 
These efforts lead to a collection of over 3.9 million audio-text pairs, with over 13,000 hours of audio.

For our neural architecture and training strategy, we propose to use a two-stage approach. 
The first stage focuses on training spectrogram-based audio encoder using a masked autoencoder (MAE) objective~\cite{he2022masked,huang2022masked}, which learns representations through masking random patches from the input spectrogram and then reconstructing these masked patches.
We anticipate that the MAE training will provide a better initialization for the following contrastive training.
An audio classification objective, in contrast, may encourage the model to discard information unnecessary for classification but important for contrastive training. 
In the second stage, we use the audio encoder from the first stage to train our audio-text model on collected synthetic audio-text pairs, employing dual contrastive and captioning objectives. 
The integration of the auxillary captioning objective, inspired by contrastive captioner (CoCa)~\cite{yu2022coca} and bootstrapping language-image pre-training (BLIP)~\cite{li2022blip}, provides stronger supervision, encouraging the audio encoder to capture fine-grained patterns that closely match text descriptions. 
Training a captioner decoder also facilitates text generation for open-ended audio understanding tasks, expanding our model's application scope.

In the evaluation phase, incorporating a diverse range of evaluation tasks is crucial to comprehensively measuring model capability and preventing overfitting to common test sets, as suggested by Recht et al.~\cite{recht2018cifar}. 
Typically, audio-text representation learning is assessed through zero-shot audio classification and audio-text retrieval. 
To provide a more comprehensive benchmark, we additionally evaluate on audio question answering (AQA)~\cite{lipping2022clotho}.
To evaluate the effectiveness of our audio encoder, we test on Holistic Evaluation of Audio Representations
(HEAR)~\cite{turian2022hear}. 
HEAR offers broad evaluation tasks that test the general-purpose audio representation through audio classification and sound event detection. 
Lastly, to assess the performance of our captioning decoder, we evaluate on automatic captioning tasks for open-ended generation.

In summary, the contribution of our work includes: (1) We curate a large-scale refined audio-text dataset with LLM processing and audio captioning.
(2) We propose a two-stage training approach for contrastive models: we first train an audio encoder with an MAE objective.
In the second stage, we train a constrative model, initializing the audio encoder from the first stage, and include an auxiliary captioning objective to enhance the model's understanding of audio-text relationships.
(3) We benchmark our model on a variety of audio understanding tasks. 
Particularly, our model achieves state-of-the-art or comparable performance on audio-text retrieval tasks. 
We have also conducted comprehensive ablation studies to demonstrate the impact of our different contributions.
We open source the inference and evaluation codebase along with our pretrained model\footnote{https://github.com/gzhu06/Cacophony}.

\section{Dataset Collection}
\label{sec:data}

\begin{figure*}[!t]
\centering
\includegraphics[width=0.95\textwidth]{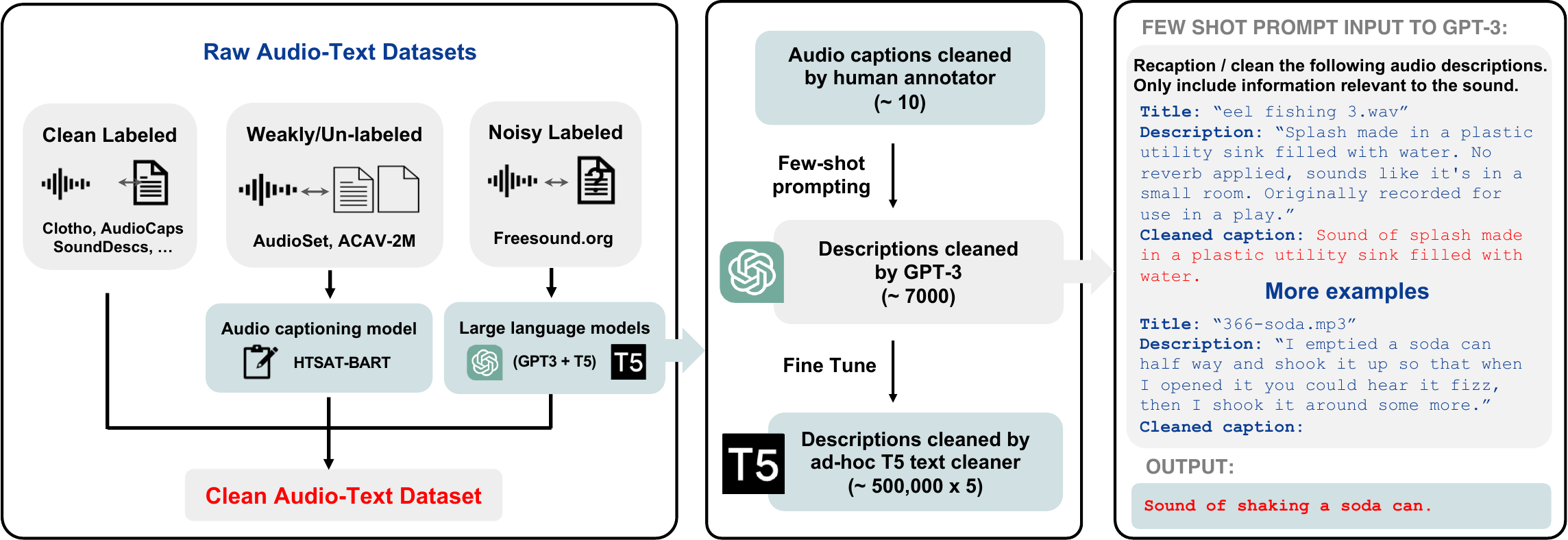}
\caption{Overview of our proposed dataset creation pipeline. 
Left: We process text descriptions based on the dataset quality. 
It aims to automatically clean and generate audio captions while maintaining consistency between the audio content and the textual descriptions.
Middle: For datasets with raw, noisy descriptions, we use large language models, GPT-3 and T5-XXL, to clean the information that is irrelevant to the sound.
Right: The detailed process of description cleaning using GPT-3 based on few-shot prompting is outlined as follows: Freesound raw inputs are highlighted in blue, sample inputs provided by a human annotator are marked in red, and the output text generated by GPT-3 is shown in bold red.}
\label{fig:data}
\end{figure*}

To make use of the audio data with either noisy or missing labels, we leverage publicly available tools to synthesize audio descriptions, shown in the left of Fig.~\ref{fig:data}.
In the following, we propose strategies to effectively handle each case.

\subsection{Clean-labeled datasets}
The ``clean-labeled'' dataset split, which we refer to as ``OpenSFX'', comprises AudioCaps-train, Clotho-development, Epidemic Sound, SoundDescs, Free To Use Sounds, Sonniss Game Effects, AudioStock, and MACS.
The most widely used datasets in this category are AudioCaps~\cite{kim2019audiocaps} and Clotho~\cite{drossos2020clotho}.
AudioCaps contains approximately 50,000 audio clips sourced from AudioSet~\cite{gemmeke2017audio} and annotated by humans. 
Clotho includes around 6,000 audio clips from Freesound\footnote{www.freesound.org}, with each clip featuring five human-generated captions.
In addition to the commonly referenced datasets, there are others like the SoundDescs dataset~\cite{koepke2022audio}, WavText5K\cite{deshmukh2022wavtext5k}, and Epidemic Sound\footnote{www.epidemicsound.com}.    
The ``writing styles" of the captions differ across audio-text datasets.
Nevertheless, 
their raw descriptions all offer detailed content information about the audio clips.
In our case, aiming to incorporate as much data as possible, we opt not to implement text-cleaning steps for processing these captions.

\subsection{Noisy-labeled datasets} 
\label{sec:noisydata}
We also include a larger-scale audio data source, Freesound, which contains noisy text descriptions. 
Freesound is a collaborative online platform dedicated to sharing sounds, hosting over 500k audio clips uploaded by users, and has been used in studies such as WavCaps~\cite{mei2023wavcaps} and LAION-CLAP~\cite{laionclap2023}.
These clips, varying in duration, cover a wide range of audio content including music, environmental sounds, synthesized effects, and various noises.
We exclude clips that are under one second or exceed five minutes, as shorter clips typically offer limited meaningful content and require excessive padding for training, and longer clips are often redundant.
While Freesound prompts data uploaders give a brief description for each audio clip, these descriptions are often inaccurate and sometimes include named entities such as people's names, locations, and details about recording equipment.

We propose to use LLMs to transform raw Freesound descriptions into usable captions by automatically removing sound-irrelevant information.
Although ChatGPT is well-suited for this task, the cost of generating captions for each Freesound sample would be prohibitively expensive. Given our resources, it would be more feasible to use a T5 model ~\cite{raffel2020exploring}, but we find that available pretrained T5 models struggle with this task. 
We thus use GPT-3 to construct a small dataset of raw-clean pairs\footnote{Model name: gpt-3.5-turbo-0301}, which will then be used to fine-tune T5-XXL~\cite{raffel2020exploring} specifically for caption cleaning. This fine-tuned T5 model is subsequently used to transform the raw Freesound descriptions to clean captions for the entire Freesound dataset, as illustrated in the middle section of Fig.~\ref{fig:data}.

More specifically, to generate the synthetic dataset used for fine-tuning T5, we few-shot prompt GPT-3 to generate a clean caption given a noisy one with 10 human annotated pairs of noisy/clean captions in our prompt. This is shown in the right section of Fig.~\ref{fig:data}.
We take a random sample of 7,000 (out of the total 500,000) Freesound noisy text captions and use GPT-3 to generate corresponding clean captions.

After fine-tuning the T5-XXL model with these pairs, we cleaned the text description for each audio sample in Freesound and generate five clean version; this servers as a form of data augmentation.

Table~\ref{tab:llm_eg} presents examples of the Freesound raw descriptions and LLM-processed captions.
Our fine-tuned T5 model is capable of automatically translating Spanish to English, eliminating redundant or irrelevant details, and summarizing long sentences into one-sentence high-level audio captions.
However, the model inaccurately represents the audio recordings in a small portion of the captions.
For instance, some captions mistakenly identify `bike' as `motorbike' or fail to entirely remove entities, as exemplified at the bottom of Table~\ref{tab:llm_eg}.

\begin{table}[!t]
\caption{Examples of raw descriptions and cleaned descriptions from our fine-tuned T5-XXL model.
Blue fonts represent successful examples, while red fonts represent failure modes.}
\centering
\resizebox{0.95\columnwidth}{!}{
\begin{tabular}{c|c}
\toprule
\textbf{Freesound raw description input} &\textbf{LLM filtered output} \\
\hline
 \makecell{$\text{Title: 06\_beso.wav}$\\ Description: sonido de beso} & \color{blue}Sound of a kiss. \\
\hline
 \makecell{$\text{Title: nice wave.WAV}$ \\Description: "i was fiddling around with an analog \\synth that went into a peavey km50 keyboard amp \\and i made this sound it has loads of pitch editing \\giving it the house-ey feel to it and i think it was \\an fmaj5/9 chord with extra 6ths and 4ths and this \\one is just a bit longer than the other one"} &\color{blue}\makecell{Analog synth sound with pitch editing \\creating a house-ey feel. Fmaj5/9 chord \\with additions of 6ths and 4ths.}\\
 \hline
\makecell{$\text{Title:Bike04.wav}$\\Description:recorded with H4N.}&\color{red}Sound of a motorbike. \\
 \hline
\makecell{$\text{Title: "Temporal process.mp3"}$\\ Description: Dr. Wineski of Morehouse school of \\Medicine reading a list of anatomical vocabulary\\ words aloud.}&\color{red}\makecell{Dr. Wineski reading a list of \\anatomic vocabulary words aloud.}\\
\bottomrule
\end{tabular}}
\label{tab:llm_eg}
\end{table}

\subsection{Weakly-labeled or unlabeled datasets} 
\label{sec:weakdata}
Lastly, the datasets with the largest scale are those that are weakly-labeled or unlabeled, including AudioSet~\cite{gemmeke2017audio} and ACAV100M~\cite{lee2021acav100m}. 
Both AudioSet and ACAV100M are comprised of audio clips extracted from YouTube videos, offering a vast range of audio data.
AudioSet is a weakly-labeled dataset with 527 predefined sound classes of 2 million 10-second recordings, where each audio clip is marked only by the presence of sound event tags, without detailed descriptions.
ACAV100M consists of 100 million videos that exhibit high audio-visual correspondence, yet it lacks of any form of labeling.
The ACAV dataset also contains a variety of scales, ranging from 20K to 100M. 
In this study, we concentrate on ACAV2M, leaving the exploration of larger scales for future work.
To leverage these datasets and to accurately represent audio content, we employ an off-the-shelf audio captioning model HTSAT-BART pretrained on WavCaps and fine-tuned on AudioCaps~\footnote{https://github.com/XinhaoMei/WavCaps}. 
Specifically, we choose this model because it achieves state-of-the-art performance on audio captioning tasks and is trained on a diverse range of data.
Employing the chosen captioning model for synthetic caption generation differs from previous work such as BLAT~\cite{xu2023blat}, whose captioner is initially trained on the smaller-scale AudioCaps dataset.

\section{Neural Architecture}
\label{sec:sys}
\begin{figure}[!t]
\centering
\includegraphics[width=0.9\columnwidth]{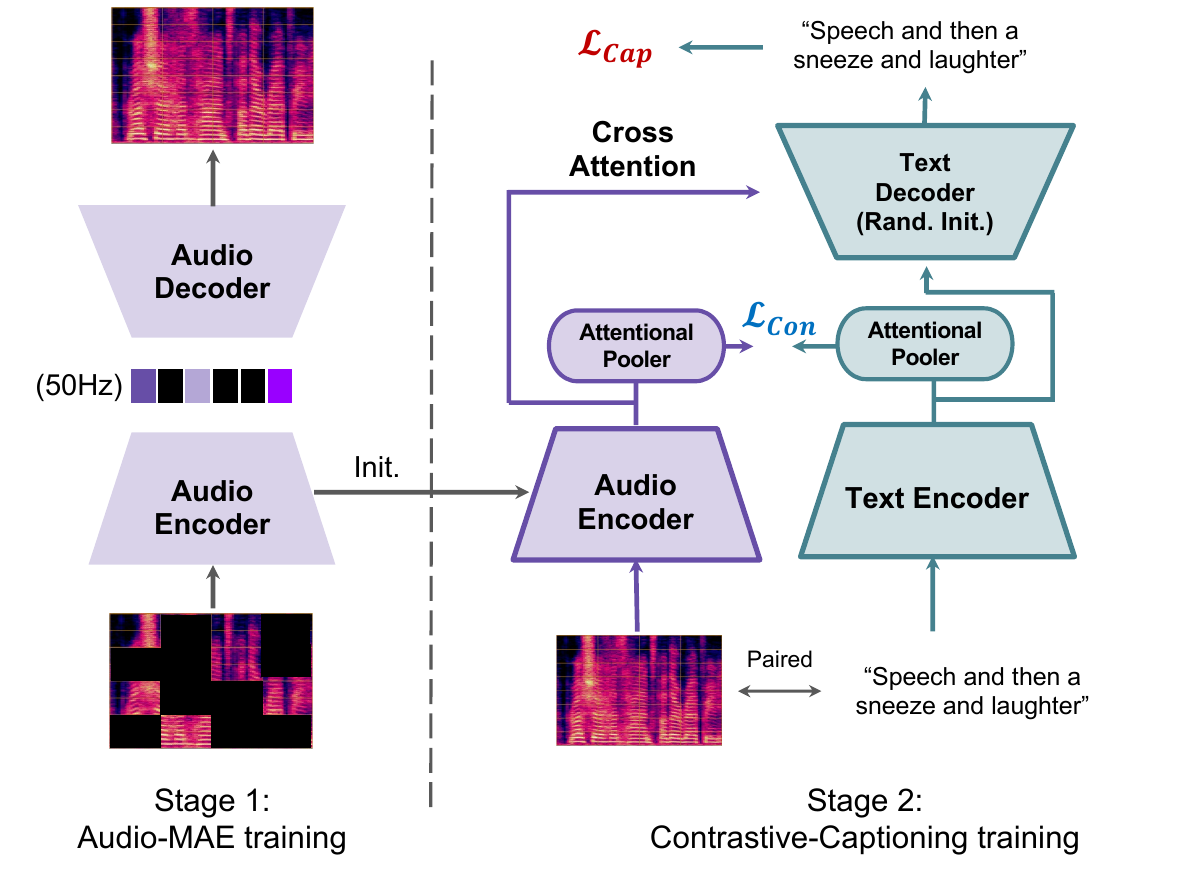}
\caption{The proposed system consists of a two-stage training process, as illustrated in the system diagram: 
Left (Stage 1): Audio-MAE training: This stage is conducted on our collected dataset.
After this stage, the audio decoder is discarded, and the encoder is retained to initialize the audio encoder for the second stage of training.
Right (Stage 2): Contrastive-captioning training: This stage involves training three components – the audio encoder, the text encoder, and the text decoder.
The second stage is dedicated to achieving a contrastive-captioning objective, aligning and fine-tuning the interaction between the audio and text components.}
\label{fig:sys_block}
\end{figure}
We train Cacophony in two stages, as depicted in Fig.~\ref{fig:sys_block}.
First, we train an audio encoder with the MAE objective, using audio-only data from our collected large-scale dataset.
Second, we take this trained audio encoder and use it as the initilzation for training the audio-text model with both contrastive and captioning objectives.

\subsection{Stage 1: Audio encoder training}
\label{sec:stage1}

We train our audio encoder with an MAE objective, which involves masking random patches of the input signals (e.g., images or spectrograms) and then reconstructing these masked portions~\cite{he2022masked,huang2022masked}. 
Specifically, in the MAE training, a transformer encoder initially processes the unmasked patches.
Following this, a transformer decoder, which will be discarded before the second-stage contrastive-captioning training, predicts both the masked and unmasked regions.
This MAE objective encourages the learning of global, contextualized representations over arbitrary subsets of spectrogram patches~\cite{huang2022masked}. 
Compared to methods based on supervised classification objectives, MAE does not required labeled data. 
It also demonstrates improved performance with increased model size~\cite{he2022masked} and increased dataset size~\cite{singh2023effectiveness}.

We initially segment the mel-spectrograms into non-overlapping regular grid patches, following Huang \textit{et al} ~\cite{huang2022masked}, who observe that utilizing overlapping patches leads to worse performance given a fixed compute budget. Subsequently, these patches are flattened, transformed via a learned linear projection, and added with positional embeddings to preserve their time-frequency ``spatial'' context.
Specifically, we employ 1-D fixed sinusoidal positional embeddings along the time axis and learnable positional embeddings along the frequency axis.
Then, 80$\%$ of the spectrogram patches are randomly masked and discarded prior to encoding. 
We chose this high masking ratio based on AudioMAE~\cite{huang2022masked}, which finds that masking 80$\%$ unstructured patches achieves the highest performance on downstream classification tasks.
The unmasked spectrogram patches are then processed by the transformer encoder. 
To obtain an input sequence for the decoder, the encoder output is concatenated sequence-wise with learnable embeddings that represent the masked patches.
Since AudioMAE~\cite{huang2022masked} demonstrates that there is no substantial performance improvement using local attention over global attention in the decoder, we maintain the original vision transformer (ViT) backbone for simplicity.
Eventually, a final linear projection layer is used to reconstruct the spectrogram patches.
Following~\cite{he2022masked,huang2022masked}, the training objective is to minimize the Mean Squared Error (MSE) between the per-patch normalized values of the reconstructed and the original input spectrogram patches.
Our preliminary experiments also show that using per-patch normalization achieves better performance in downstream tasks.
Finally, we discard the decoder and only keep the trained encoder for the second-stage training.


We elect not to use the Swin-Transformer~\cite{liu2021swin} used by LAION-CLAP~\cite{laionclap2023}, since pyramidal ViTs introduce patch merging as well as operations within ``local'' windows, which would make it difficult to directly handle the random sequence of partial spectrogram tokens~\cite{li2022uniform}. 
Aside from the issue of incompatibility with MAE, the patch-merging in Swin-Transformer results in significant time decimation, with an output resolution of 6 Hz compared to 50 Hz output of ViT. 
By avoiding decimation in the time dimension, we can readily adapt our pretrained audio encoder for tasks such as sound event detection with high time-resolution.
\vspace{-0.2cm}
\subsection{Stage 2: Contrastive-captioning training}
\label{sec:stage2}
Given the audio encoder trained in the first stage, we now introduce a text encoder/decoder and train a contrastive model with an auxiliary captioining objective, in a setup similar to BLIP~\cite{li2022blip} and CoCa~\cite{yu2022coca}. 
For the contrastive objective~\cite{jia2021scaling,radford2021learning}, we use learned linear projections to map the text and audio embeddings to the same dimension, followed by an $l_2$-normalization layer.
Then, we use matched audio-text embeddings as positives and all non-paired examples as hard negatives, resulting in $N$ total samples for the pairwise Information Noise-Contrastive Estimation (InfoNCE) loss as described in~\cite{oord2018representation}:
\begin{equation}
    \begin{multlined}
   \mathcal{L}_{\text{Con}} = -\frac{1}{N}(
    \underbrace{\sum_i\log{\frac{\exp(x_i^\top y_i / \tau)}{\sum_{j=1} \exp(x_i^\top y_j / \tau )}}}_\text{audio-to-text}\\
    +\underbrace{\sum_i\log{\frac{\exp(y_i^\top x_i / \tau)}{\sum_{j=1} \exp(y_i^\top x_j / \tau)}}}_\text{text-to-audio}),
    \end{multlined}
    \label{eq:con}
\end{equation}
where $x_i$ and $y_j$ are $l_2$-normalized embeddings of the audio in the $i$-th pair and text in the $j$-th pair, respectively, and $\tau$ is a learnable temperature.

For the captioning objective, the model is required to autoregressively predict the tokenized text associated with a given audio sample.
The text decoder is trained to minimize the negative log-likelihood of current ground-truth token given previous ground-truth tokens:
\begin{equation}
\mathcal{L}_{\text{Cap}}=-\frac{1}{T}\sum_{t=1}^T\log P_{\theta}(y_t|y_{1:t-1},x),
\end{equation}
where $y_t$ is the $t$-th ground-truth token for a given caption $y$ and $T$ is the caption's total length.
As a result, we apply both contrastive and generative objectives in the second stage training as follows:
\begin{equation}
    \mathcal{L}_{\text{II}} = \mathcal{L}_{\text{Con}} + \lambda_{\text{Cap}} \cdot \mathcal{L}_{\text{Cap}},
    \label{eq:ii}
\end{equation}
where $\lambda_{\text{Cap}} > 0$ is a hyperparameter controlling the relative weight of the captioning loss compared to the contrastive loss.

Architecturally, the second-stage training involves training three key modules: audio encoder, text encoder, and text decoder.
For the audio encoder, we use the encoder from the first stage of AudioMAE training, outlined in Sec.~\ref{sec:stage1}. 
Our text encoder is a transformer that follows RoBERTa's architecture. 
However, unlike RoBERTa, we apply causal self-attention rather than bidirectional self-attention. 
This is a necessary modification to prevent information leakage to the text decoder.
We initialize our model with the RoBERTa pretrained weights.
The outputs from audio encoder and text encoder are framewise and token-wise embeddings respectively; in order to obtain a single embedding vector for each modality, we choose to integrate multi-head attention poolers on top of the sequential embeddings~\cite{lee2019set,yu2022coca} for the contrastive objective, shown in the right of Fig.~\ref{fig:sys_block}.
Our text decoder consists of stacked transformer layers on top of our text encoder with causal self-attention to prevent next-token prediction leakage.
Each transformer block has a cross-attention layer that attends to the full output of our audio encoder.  
The text decoder layers are initialized randomly.

Because our audio data used for training varies in length, we follow a similar training strategy as the first MAE training stage: for audio shorter than the training length, we utilize zero-padding and generate corresponding masks;
for audio exceeding the training length, we randomly sample time-frequency patches from the full spectrograms and feed them into the audio encoder. 
It is worth noting that for shorter audio, this differs from the first-stage training: During MAE training, 80$\%$ of spectrogram patches are masked, whereas during contrastive training, audio shorter than the predefined training length is fully unmasked.

\section{Implementation Details}
\label{sec:train}
\subsection{Training and inference setup}
As outlined in Sec.~\ref{sec:data}, the training datasets are categorized into three types: the `clean-labeled' dataset consists of 1,212 hours, the `noisy labeled' dataset consists of 3,003 hours, and the `weakly/unlabeled' dataset consists of 9,017 hours. 
All audio files are processed into a mono channel with a sampling rate of 16 kHz.
We extract mel-spectrograms using a 25 ms window and a 10 ms hop length, with 128 mel bands and 512 as FFT size.
The spectrograms are then converted into 16$\times$16 patches without overlap. 

In the first-stage training, we randomly sample 15 seconds of audio for each recording, corresponding to a patch sequence length of 750.
At the second stage of training, we limit the audio length to 10.24 seconds, equivalent to a patch sequence length of 512, i.e., when the patch sequence length exceeds 512, we randomly sample 512 patches from the full sequence, following Patchout faSt Spectrogram Transformer (PaSST)~\cite{koutini2021efficient}.
Text samples are tokenized with the pretrained RoBERTa tokenizer and truncated to maximum length of 77.
Our audio encoder, employed in both training stages, is a 12-layer ViT-Base (ViT-B) Transformer with 8 attention heads, a hidden size of 768 and an intermediate size of 3072, and utilizes the SiLU activation function. 
Additionally, a 12-layer decoder with the same settings is used during the MAE training phase.
For the attention pooler, we choose to use $n_{head}=8$.

When training AudioMAE, we use a batch size of 512 and a learning rate of $2\times10^{-4}$, and employ the Adam with decoupled Weight decay (AdamW)~\cite{loshchilov2017decoupled} optimizer with a weight decay of 0.01. 
We train for 200,000 steps, with a 10,000 step learning rate warm-up and a cosine decay to $1\times10^{-6}$.
In the contrastive stage, we set $\lambda_{\text{Cap}} = 1$ as the default value. 
While fine-tuning this parameter may marginally improve performance, as suggested by Yu et al.~\cite{yu2022coca}, we find the default setting to be sufficiently effective for our purposes.
During training, we again use AdamW with weight decay 0.01 with a batch size of 4096. 
The learning rate is warmed up for 10,000 to a peak of $1\times10^{-5}$ and is decayed to $1\times10^{-6}$ over 300,000 (total) steps following a cosine schedule.

During inference, we process the full audio sequence if it fits in the memory. 
For longer sequences exceeding memory capacity, we randomly sample a subset of spectrogram patches that fit within available memory. 
This approach allows us to handle variable-length inputs while adapting to computational limitations.
Notably, despite the potential for duration distribution shift between training (where we use masked or dropped spectrogram patches) and inference (where we process full sequences), we observe improved performance when utilizing all available audio information during inference, aligning with findings in~\cite{li2023scaling}.

\vspace{-0.2cm}
\subsection{Sharpness-aware minimization}
Given our use of a large training batch size of 4096 for the contrastive objective and only having nearly 4 million pairs of audio-text data, one epoch is completed in only around 1,000 steps.
When training our contrastive model initially, we empirically observe overfitting early in the training process.
DALL·E-2~\cite{ramesh2022hierarchical} found that using Sharpness-Aware Minimization (SAM)~\cite{foret2020sharpness} in training CLIP models improves performance; we thus investigate using SAM in our contrastive training stage. 
SAM is an optimization technique that encourages convergence toward flatter local minima with the hope of improving model generalizability, which is also shown to provide robustness against label noise~\cite{foret2020sharpness}.

More concretely, we denote $S= {(\boldsymbol{x}_i, y_i)}^n_{i=1}$ as the training dataset and $\boldsymbol{w}$ as the trainable parameters.
SAM seeks to find the parameters whose neighbors have low training loss through the following objective:
\begin{equation}
    \min_{\boldsymbol{w}} \mathcal{L}_{S}^{SAM}(\boldsymbol{w})= \min_{\boldsymbol{w}} \max_{\|\boldsymbol{\epsilon}\|_{p}\leq \rho} \mathcal{L}_{S}(\boldsymbol{w}+\boldsymbol{\epsilon}
    ),
\label{loss_sam}
\end{equation}
where $p \geq 1$ defines the order of the norm and $\rho\geq 0$ is the tunable hyperparameter that measures the size of the neighborhood, and $\mathcal{L}_S$ can be any arbitrary loss function on dataset $S$.
The inner maximum operator computes the maximum loss within the neighborhood. Its difference from $\mathcal{L}_S(\boldsymbol{w})$ defines how quickly the loss increases from $\boldsymbol{w}$, i.e., the sharpness of the loss landscape. The SAM loss is thus minimizing not only the $\mathcal{L}_S(\boldsymbol{w})$ itself, but also the sharpness within the neighborhood.
To solve the inner maximization problem efficiently, we can use the first-order Taylor expansion approximation, i.e., $\mathcal{L}_{S}(\boldsymbol{w}+\boldsymbol{\epsilon}) \approx \mathcal{L}_{S}(\boldsymbol{w})+\boldsymbol{\epsilon}^T\nabla_{\boldsymbol{w}}\mathcal{L}_{S}(\boldsymbol{w})$
~\cite{foret2020sharpness}. 
With $p=2$, the inner maximization can be achieved at:
\begin{equation}
    \boldsymbol{\hat{\epsilon}}(\boldsymbol{w}) \approx \arg\max_{\|\boldsymbol{\epsilon}\|_{2}\leq \rho} \epsilon^T\nabla_{\boldsymbol{w}}\mathcal{L}_{S}(\boldsymbol{w})=\rho\frac{\nabla_{\boldsymbol{w}}\mathcal{L}_{S}(\boldsymbol{w})}{||\nabla_{\boldsymbol{w}}\mathcal{L}_{S}(\boldsymbol{w})||}, 
\label{eq:sam_grad}
\end{equation}
which is a scaled version of the gradient, normalized to have a magnitude of $\rho$.
Substituting $\boldsymbol{\hat{\epsilon}}$ back into equation \eqref{loss_sam} and differentiating it w.r.t. $\boldsymbol{w}$, we obtain: 
\begin{equation}
\nabla_{\boldsymbol{w}}\mathcal{L}_{S}^{SAM}(\boldsymbol{w})\approx \nabla_{\boldsymbol{w}}\mathcal{L}_{S}(\boldsymbol{w}+\boldsymbol{\hat{\epsilon}}(\boldsymbol{w}))\approx \nabla_{\boldsymbol{w}}\mathcal{L}_{S}(\boldsymbol{w})|_{(\boldsymbol{w}+\boldsymbol{\hat{\epsilon}})},
\label{eq:sam_calculate}
\end{equation}
where the first approximation is due to the Taylor expansion mentioned before, while the second approximation is due the removal of a higher-order differentiation term.
Note that two forward/backward passes are required to compute every update: one for computing the perturbation $\hat{\boldsymbol{\epsilon}}$ and the other one for computing the gradient at the perturbed weight vector $\boldsymbol{w} + \hat{\boldsymbol{\epsilon}}$; thus, the wall-clock time per step is essentially doubled. 
There has been recent work~\cite{liu2022towards} in reducing SAM's computational overhead; we leave the application of these techniques to future work.

\begin{figure}[!t]
\centering
\includegraphics[width=0.7\columnwidth]{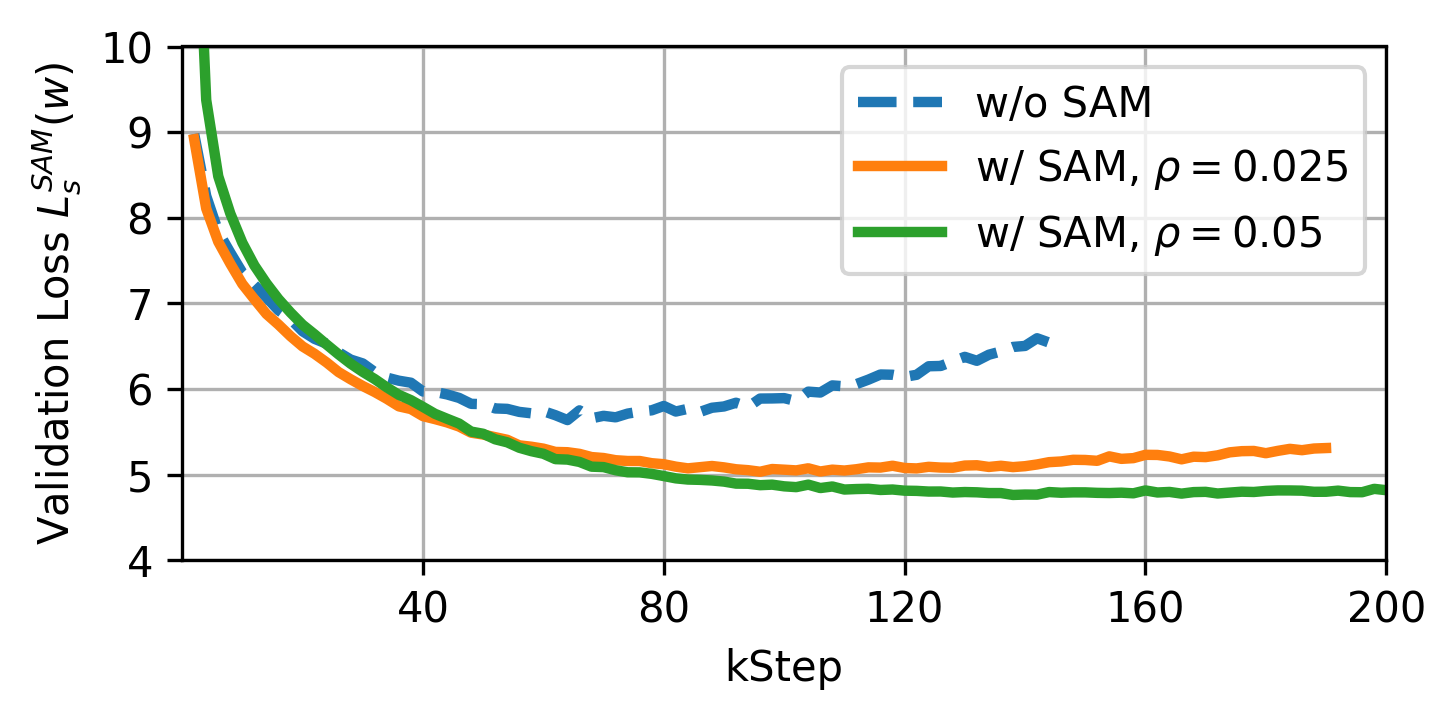}
\caption{Contrastive-Captioning objective on the validation set during training, comparing scenarios with and without the application of SAM. 
For the cases with SAM optimization, we employ various neighborhood sizes as determined by the hyperparameter $\rho$.
}
\label{fig:sam_loss}
\end{figure}

The training dynamics, depicted in Fig.~\ref{fig:sam_loss}, show the impact of incrementally adjusting $\rho$ by increments of 0.025.
Increasing $\rho$ corresponds to a larger $\boldsymbol{\epsilon}$ perturbation and thus heavier regularization. The figure clearly shows that without SAM, the model is prone to overfitting on the training data as early as 40,000 steps.
In contrast, with even a small degree of sharpness minimization at $\rho=0.025$, overfitting is delayed until approximately 100,000 steps and occurs at a significantly lower objective value.
In our final second-stage model training, we use SAM with $\rho=0.075$.


\section{Evaluation}
\subsection{Evaluation overview} 
By employing both contrastive and captioning training objectives, Cacophony is capable of providing audio-text representations and generating free-form text for open-ended audio understanding tasks, whereas the vanilla CLAP models cannot generate text without separate adapter modules.

\subsubsection{Tasks}
We evaluate Cacophony's performance across three primary categories: (1) Audio-Text Representation Tasks: These include audio-text retrieval, closed-ended audio question answering, and zero-shot transfer in audio classification. 
These assessments concentrate on the model's performance in audio-text cross-modal alignment.
(2) General-Purpose Audio Representation Tasks: To evaluate Cacophony's pretrained audio encoder as a general-purpose audio representation model, we conduct audio classification and sound event detection tasks. 
For these assessments, we train a Multi-Layer Perceptron (MLP) on top of our frozen audio encoder.
(3) Open-Ended Audio Understanding Task: We assess our model's capacity of generating free-form text through audio captioning. 
However, we have excluded open-ended audio question answering from our evaluation, as it requires additional text pre-processing and fine-tuning steps beyond the scope of this study.
\subsubsection{Evaluation datasets}
In the audio research community, some major high-impact datasets originate from Freesound, including Clotho~\cite{drossos2020clotho}, ESC50~\cite{piczak2015dataset}, UrbanSound8K~\cite{Salamon14UrbanSound}, and FSD50K~\cite{fonseca2021fsd50k}. 
Another set of popular datasets comes from YouTube, such as Audioset (and its subset AudioCaps) and VGGSound~\cite{chen2020vggsound}.
Given that our dataset includes Freesound and 4 million Youtube samples, we need to ensure that none of the data on which we evaluate is present in our train dataset.
FSD50K, a subset of Freesound, is originally in the HEAR benchmark. Because FSD50K is of high-quality and is relatively large, rather than removing FSD50K from our training dataset, we exclude it from our HEAR benchmark evaluation.
\textit{Mridingham Stroke and Mridingham Tonic} also overlaps with Freesound; we choose to exclude it from the HEAR benchmark for simplicity, as it is small-scale relative to other HEAR tasks. Lastly, we exclude \textit{Beehive} tasks from our HEAR evaluation because they require inference on 600-second long samples, and the models to which we compare are unable to process audio of this length.

\begin{center}
    \begin{table*}[!t]
  \caption{System comparisons on audio-language retrieval on test sets of AudioCaps and Clotho. Results for the baselines are copied from the references. $R@k$ (Recall at $k$) represents the proportion of queries for which the relevant item is retrieved within the top $k$ results. Higher values indicate better performance. 
  }
  \label{tab:retrieval_results}
  \centering
  \small
  \resizebox{0.85\textwidth}{!}{
  \begin{tabular}{c|ccc|ccc|ccc|ccc}
    \toprule
    \multirow{3}{*}{\textbf{Model}} & \multicolumn{6}{c|}{\textbf{AudioCaps}} & \multicolumn{6}{c}{\textbf{Clotho}} \\
    \cline{2-13}  & 
    \multicolumn{3}{c|}{\textbf{Text-to-Audio}} & \multicolumn{3}{c|}{\textbf{Audio-to-Text}} & \multicolumn{3}{c|}{\textbf{Text-to-Audio}} & \multicolumn{3}{c}{\textbf{Audio-to-Text}} \\
    \cline{2-13}
    &  $\boldsymbol{R@1}$ & $\boldsymbol{R@5}$ & $\boldsymbol{R@10}$ & $\boldsymbol{R@1}$ & $\boldsymbol{R@5}$ & $\boldsymbol{R@10}$ & $\boldsymbol{R@1}$ & $\boldsymbol{R@5}$ & $\boldsymbol{R@10}$ & $\boldsymbol{R@1}$ & $\boldsymbol{R@5}$ & $\boldsymbol{R@10}$ \\
    \hline 
    CLAP-HTSAT \cite{deshmukh2022wavtext5k} & 34.6 & 70.2 & 82.0 & 41.9 & 73.1 & 84.6 & 16.7 & 41.1 & 54.1 & 20.0 & 44.9 & 58.7 \\
    LAION  \cite{laionclap2023} & 36.1 & 71.8 & 83.9 & 46.8 & 82.9 & 90.7 & 16.1 & 38.3 & 51.1 & 22.7 & 48.5 & 60.8 \\
    LAION (fusion) \cite{laionclap2023} & 35.1 & 71.5 & 83.6 & 45.8 & 80.9 & 91.6 & 18.2 & 42.5 & 54.4 & 25.7 & 51.5 & 63.4 \\
    WavCaps-CNN14~\cite{mei2023wavcaps} & 34.7 & 69.1 & 82.5 & 44.6 & 76.3 & 86.2 & \textbf{21.2} & 46.4 & \textbf{59.4} & 25.9 & 52.6 & 65.8 \\
    WavCaps-HTSAT~\cite{mei2023wavcaps} & 39.7 & 74.5 & 86.1 & 51.7 & 82.3 & 90.6 & 19.5 & 45.2 & 58.2 & 23.4 & 50.9 & 63.4 \\
    FLAP~\cite{yeh2023flap}&40.4&74.7&85.0&51.5&82.5&92.5&17.4&41.3&53.7&21.6&51.2&63.1\\
    FLAP (fusion)~\cite{yeh2023flap}&\textbf{41.5}&\textbf{75.5}&86.0&53.0&\textbf{84.1}&\textbf{92.6} &20.3&\textbf{46.5}&58.8&25.5&53.4&\textbf{67.9}\\
    Cacophony (ours)& 41.0 & 75.3 & \textbf{86.4} & \textbf{55.3} & 83.6 & 92.4 & 20.2 & 45.9 & 58.8 & \textbf{26.5} & \textbf{54.1} & 67.3 \\
    \bottomrule
  \end{tabular}
  }

\end{table*}
\end{center}
\vspace{-1cm}
\subsection{Audio-language retrieval}
\label{sec:atr}
\subsubsection{Task definition}
The audio-text retrieval task involves searching for a specific audio clip or a caption based on a query from the other modality.
Text-to-audio retrieval involves retrieving audio for a given text caption, and audio-to-text retrieval involves retrieving text for a given audio sample.

\subsubsection{Experimental setup}
To effectively perform retrieval tasks, pretrained contrastive models are used to predict if a given audio clip and text description are paired together. 
Following prior works~\cite{radford2021learning,laionclap2023,elizalde2022clap}, for audio-to-text retrieval, we first compute the feature embeddings for the target audio clip and the corresponding set of potential text captions.
Subsequently, we compute the cosine similarity between the audio embedding and every text embedding.
The retrieved text samples are then selected based on the highest cosine similarity scores (the text-to-audio retrieval process is identical with the roles reversed).
During evaluation, we benchmark the audio-text retrieval task on the test splits of AudioCaps and Clotho datasets. 
We use recall at rank $k$ ($R@k$) as our evaluation metric.
For a given query, $R@k$ is assigned a value of 1 if the relevant item is among the top $k$ retrieved items and 0 if it is not.
This $R@k$ score is then averaged across the entire dataset to obtain the final performance.
\subsubsection{Result}
The audio-language retrieval results on the AudioCaps and Clotho datasets are presented in Table~\ref{tab:retrieval_results}, where we compare against the most recent contrastive-based models, including MS-CLAP~\cite{deshmukh2022wavtext5k}, LAION-CLAP~\cite{laionclap2023}, WavCaps~\cite{mei2023wavcaps} and FLAP~\cite{yeh2023flap}.
To ensure a fair comparison, we only compare with pretrained models, rather than models that are fine-tuned for specific audio-text retrieval tasks.

Our proposed method has achieved state-of-the-art or comparable performance to the best-performing systems across all evaluated metrics and both datasets. 
Compared to WavCaps and LAION, our model achieves better performance on AudioCaps and comparable results on Clotho with `WavCaps-CNN14'. 
When compared against FLAP, Cacophony outperforms the non-feature-fusion-based model and matches the fusion-based model.
While both Cacophony and FLAP share some common elements, such as LLM-augmented captions and MAE-based audio encoders, our model's superior performance can be attributed to our two-stage MAE and contrastive training strategy, together with SAM optimization techniques.
This combination allows for more effective learning of audio-text representations, resulting in improved retrieval performance across both datasets.

\begin{center}
    \begin{table}[t]
\caption{Evaluation of audio question answering in Clotho-AQA and Music-AQA benchmarks (``Word" stands for single-word closed-vocabulary classification, ``Bin.'' stands for binary classification. ``Acc.'' stands for accuracy)}
\label{tab:aqa}
\centering
\resizebox{\columnwidth}{!}
{
\begin{tabular}{cccc|c|c|c}
\toprule
\multirow{2}{*}{Recall ($\%$)}  & \multicolumn{3}{c}{Clotho-Word} & \multicolumn{1}{c}{Clotho-Bin.} & \multicolumn{1}{c}{Music-Word} & \multicolumn{1}{c}{Music-Bin.} \\
\cline{2-7}\specialrule{0em}{1pt}{1pt}
\multicolumn{1}{c}{} & $\boldsymbol{R@1}$ & $\boldsymbol{R@5}$ & $\boldsymbol{R@10}$ & \bf{Acc.} &\bf{Acc.}& \bf{Acc.}\\
\hline 
MWAFM~\cite{li23Multi} & \bf{21.3}   & \bf{45.5}& \bf{56.7} & 68.6   & 58.0& 70.5 \\
\hline 
LAION & 17.8 & 42.1 & 53.2 & 68.4  & \bf{58.9} & 69.2\\
LAION (fusion)& 18.7& 42.4 & 53.3  & 68.2  & 57.4  & 71.2\\
MS-CLAP& 19.4& 43.7 & 54.5& 68.8    & 53.7 & 74.5\\
WavCaps-CNN14  & \textbf{20.3} & \textbf{44.3} & \textbf{55.5}    & 66.8  & 54.3 & 71.1\\
WavCaps-HTSAT & 18.1  & 41.5 & 53.0  & 68.4& 53.3& 74.2   \\
Cacophony (Ours) & 19.7 & 42.6 & 52.0  & \bf{70.7}  & 53.6  & \bf{74.9}    \\
\bottomrule
\end{tabular}
}
\end{table}
\end{center}
\vspace{-1cm}
\subsection{Closed-ended audio question answering}

\subsubsection{Task definition}
AQA is the task of generating a text response to a text question about an audio signal.
We follow the strategy in~\cite{lipping2022clotho} to cast AQA into a supervised classification task with a closed-ended answer set.

\subsubsection{Experimental setup}
We evaluate our approach on two datasets: Clotho-AQA~\cite{lipping2022clotho} and Music-AQA~\cite{li2022learning}. 
Clotho-AQA contains 1,991 audio samples, each with six questions. 
Four questions have yes/no answers, while two have single-word answers. 
Music-AQA, derived from the Music-AVQA dataset, includes 6,319 one-minute audio clips from musical performances. 
Each clip has one question assessing counting or comparison skills. 
Both datasets feature binary or single-word answers, with Music-AQA responses ranging from `zero' to `nine'.
We evaluate closed-ended AQA through training small MLPs on top of frozen pretrained encoders.
In particular, we first extract embeddings from the frozen audio and text encoders from the contrastive models, and then we concatenate them into one vector and pass it through a 4-layer MLP for binary or multi-class classification.

In addition to several audio-text models, we also compare our approach to a baseline model~\cite{li23Multi} designed specifically for AQA that achieves state-of-the-art performance on the Clotho-AQA and Music-AQA benchmarks. 
This model differs from contrastive models in two ways: First, its audio and text encoders are pretrained independently, and second, after extracting frame-level audio and token-level word embeddings, these embeddings are processed through fused attention layers followed by MLP layers.
We follow the train, validation and test splits provided in the original Clotho-AQA and Music-AQA datasets.
We use $R@k$ to evaluate the performance on the Clotho-AQA dataset because the size of the answer vocabulary is large (828). 
We use accuracy to evaluate the performance on the Music-AQA dataset because its single-word answer vocabulary consists of only 10 possible answers to different questions.

\subsubsection{Result}
In our evaluations, detailed in Table~\ref{tab:aqa}, we benchmark fine-tuned contrastive models together with a state-of-the-art AQA baseline. 
Although Cacophony outperforms other contrastive models in binary question tasks, it performs less well in single-word classification compared to the specialized AQA baseline.
We believe this is because Cacophony uses pooled embeddings from a frozen text encoder, while the baseline learns to aggregate over granular text features ~\cite{li23Multi}.
As the text in AQA task, \textit{i.e.}, a question, is out-of-distribution for our text encoder, the frozen pooled embeddings may not precisely capture the question's semantics; 
A trainable pooler could offer improved pooled embeddings by leveraging the fine-grained output of the encoder.


\begin{center}
    \begin{table}[!t]
  \caption{System comparisons on zero-shot classification. Each dataset is marked with the original sampling rate. *:Due to the inaccessibility of many YouTube videos, our VGGSound test split comprises a total of 12,722 samples.}
  \label{tab:zs}
  \centering
  \resizebox{0.9\columnwidth}{!}{
  \begin{tabular}{c|c|c|c|c}
    \toprule
    Accuracy ($\%$) & VGGSound* & TUT-AS & ESC-50 & US-8K\\
      & (48k) & (44.1k) & (44.1k) & (44.1k)\\
    \hline
    MSCLAP \cite{deshmukh2022wavtext5k}  & 12.3 & 25.1 & 81.6 & 73.1\\
    \hline
    LAION  \cite{laionclap2023} & \bf{30.0} & 47.4 & 84.4 & 76.1 \\
    LAION (fusion) \cite{laionclap2023}& 26.9 & 27.5 & 84.1 & 71.6 \\
    \hline
    WavCaps-CNN14~\cite{mei2023wavcaps} & 27.5 & 47.2 & 87.0 & 72.9 \\
    WavCaps-HTSAT~\cite{mei2023wavcaps}  & 28.9 & \bf{49.4} & 93.1 & \bf{80.4} \\
    \hline
    Cacophony (ours)& 27.1 & 48.6 & \bf{93.4} & 77.1 \\

    \bottomrule
  \end{tabular}
  }
\end{table}
\end{center}

\vspace{-1cm}
\subsection{Zero-shot transfer on audio classification}
\subsubsection{Task definition}

Audio classification is to categorize audio recordings into different sound types.
To evaluate the generalization capability of the contrastive models on unseen datasets, we explore their zero-shot transfer ability on these audio classification tasks following previous practices~\cite{radford2021learning,laionclap2023,elizalde2022clap}.
This setup resembles the audio-to-text retrieval task defined in Sec.~\ref{sec:atr}, with the key difference being the use of a predefined set of labels as text descriptions. 

\subsubsection{Experimental setup}
We evaluate the models' zero-shot classification accuracy using four datasets: VGGSound-test, TUT Acoustic Scenes~\cite{mesaros2018multi}, ESC-50~\cite{piczak2015dataset}, and UrbanSound8K~\cite{Salamon14UrbanSound}.
These datasets collectively cover a wide range of sound events. 
For each dataset during classification, we use the names of all classes within that dataset as the set of possible text descriptions, and as in audio-to-text retrieval, select the matching text based on cosine similarity.
Since the text descriptions seen during training are relatively verbose compared to the often-single-word labels in classification tasks, we craft individual prompt templates for each dataset.
For instance, in the context of sound event classification, we use a template such as ``This is a sound of [label]", while for acoustic scene classification, we use ``This sound is on [label]."
We use Top-1 accuracy as the evaluation metric.

\subsubsection{Result}
We compare our model, Cacophony, against all baseline models on zero-shot classification on a variety of audio classification benchmarks, presented in Table~\ref{tab:zs}.
Overall, Cacophony exhibits competitive performance for general sound categories on ESC-50, UrbanSound8K, and TUT-Acoustic Scenes.
However, it achieves a lower accuracy on the VGGSound dataset.
This relatively lower performance on VGGSound, which includes 310 classes, 
suggests that its granularity may be too fine for the model to differentiate different classes.
This challenge is likely due to the presence of inaccuracies within the re-captioned audio descriptions in the training data, potentially resulting from both the LLM cleaning and automatic captioning processes. 
Additionally, since our model is trained on audio sampled at 16,000 Hz, training on full-bandwidth audio recordings could potentially improve its performance in zero-shot classification tasks, as demonstrated in previous work~\cite{kong2020panns}.

\vspace{-0.2cm}
\subsection{Holistic evaluation of audio representations}

\begin{figure}[!t]
\centering
\includegraphics[width=0.75\columnwidth]{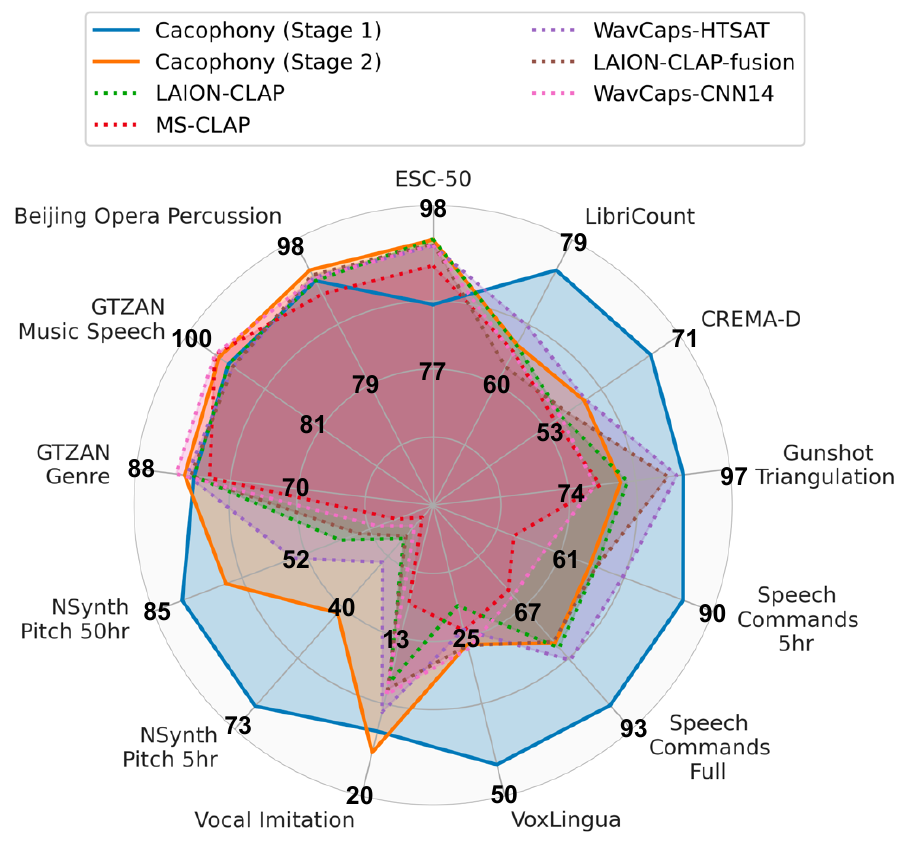}
\caption{Comparison of classification accuracy ($\%$) on HEAR benchmark.
Evaluation scores are
stable across tasks, with a median 95$\%$ confidence interval of 0.25$\%$ with \texttt{hear-eval-kit} model-selection strategy. 
Comparisons of audio branches from contrastive language-audio embeddings.}
\label{fig:conhear}
\end{figure}

\subsubsection{Task definition}
The HEAR benchmark is designed to evaluate the effectiveness of general audio representations across various domains, including speech, music and general sound. 
The HEAR benchmark comprises two principal task types: (1) scene-based tasks, which involve classifying an entire audio clip, and
(2) event-based tasks, which aim at identifying specific sound events over time, i.e., predicting the start time, end time, and label for each sound event.
\begin{center}
    \begin{table}[!t]
\caption{Comparison with top-performing non-ensemble based systems on event-based tasks in HEAR benchmark leaderboard using onset F1 score ($\%$). Results from other systems are taken directly from the HEAR Leadboard.}
\label{tab:hearsed}
\centering
\resizebox{0.75\columnwidth}{!}{
\begin{tabular}{c|c|c}
\toprule
Event Onset F-1 ($\%$) &DCASE2016 T2 &MAESTRO 5h\\
\hline
OpenL3~\cite{cramer2019look} 
&\bf{83.2}& 1.7 \\
wav2vec2~\cite{baevski2020wav2vec} 
&66.3&3.3 \\
SONY-ViT  & 66.8 &23.9 \\ 
CREPE~\cite{kim2018crepe}
&50.4 & \bf{40.1} \\
Cacophony (Ours)  & 81.1 & 10.0 \\ 
\bottomrule
\end{tabular}
}
\end{table}
\end{center}
\subsubsection{Experimental setup}
\vspace{-0.7cm}
We follow the fine-tuning strategies presented in~\cite{turian2022hear} using \texttt{hear-eval-kit}\footnote{https://github.com/hearbenchmark/hear-eval-kit}. 
Specifically, in both scene-based and event-based tasks, the audio encoders from the audio-text models are frozen and used as the input feature vector to a shallow downstream MLP classifier.
During evaluation following~\cite{turian2022hear}, scene-based tasks are measured with classification accuracy. 
Event-based tasks are evaluated with event onset F-measure, which correlates better with human perceptual than framewise classification accuracy~\cite{hawthorne2018onsets}.

\subsubsection{Result}
We evaluate our models against existing state-of-the-art contrastive models using the HEAR benchmark, shown in Fig.~\ref{fig:conhear}.
Our proposed system demonstrates good performance across various tasks, outperforming or closely matching existing approaches. 
Notably, it achieves significantly better accuracy than other contrastive models in the pitch classification tasks defined in \textit{NSynth Pitch}. 
WavCaps-HTSAT and our Cacophony models show relatively balanced performance across different datasets but no single model is superior across all tasks.
Cacophony underperforms WavCaps-HTSAT baseline in several speech-related datasets, including \textit{Speech Commands} and \textit{LibriCount}.
We believe that incorporating more speech-specific data would improve performance on these speech-related tasks. 

Interestingly, in a comparison between Cacophony (stage 1) and Cacophony (stage 2), \textit{i.e.,} before and after the contrastive-captioning training, we notice a significant drop in performance across various tasks.
This suggests that the contrastive training objective, which encourages the audio encoder to extract features relevant to selecting a text pair, may not align with other audio classification objectives.

For event-based tasks, we do not compare our model with existing contrastive models, because they tend to perform poorly without manually computing shifting-windowed embeddings, as they typically use temporal pooling in convolutional blocks or downsampling in Swin blocks~\cite{liu2021swin}.
Instead, we compare our model against top-performing individual models (i.e., excluding systems that ensemble multiple models) on the sound event detection tasks listed on the HEAR leaderboard\footnote{HEAR leaderboard (https://hearbenchmark.com/hear-leaderboard.html)}. We exclude ensemble systems for a fair comparison on individual model performance. 

Table~\ref{tab:hearsed} shows the results.
Our model demonstrates competitive performance in the ``DCASE 2016 task 2'' with a score of 81.1$\%$, which is among the top-performing systems.
However, its capability appears more limited in the ``MAESTRO 5h'' task, scoring only 10.0$\%$. 
This suggests that it may not be as effective in tasks that require fine-grained instrumental pitch detection, in contrast to daily environmental sounds. 


\begin{center}
    \begin{table}[t]
\caption{Automated audio captioning results on test sets of AudioCaps and Clotho. 
All compared baselines are from WavCaps~\cite{mei2023wavcaps}, the evaluation results for the baselines are from the reference. 
HTSAT-BART-FT is fine-tuned specifically on AudioCaps, and we also used it for recaptioning weakly/un-labeled datasets. 
}
\label{tab:caption}
\centering
\resizebox{\columnwidth}{!}{
\begin{tabular}{c|c c c c c c c}
\toprule
\textbf{Model} & BLEU$_1$ &BLEU$_4$ &ROUGE$_l$ &METEOR &CIDER &SPICE &SPIDEr\\
\midrule
\textit{AudioCaps} \\
HTSAT-BART-FT& \textbf{70.7} &\textbf{28.3} &\textbf{50.7} & \textbf{25.0} &\textbf{78.7} &\textbf{18.2} &\textbf{48.5}\\
CNN14-BART&67.0 &26.1 &48.3 &23.1 &72.1 &16.9 &44.5\\
HTSAT-BART & 67.5 &27.2 &48.3 &23.7 &71.1 &17.7 &44.4\\
Cacophony (Ours)& 68.4 &25.9 &48.6&23.6  &72.8&16.8&44.8\\
\midrule
\textit{Clotho} \\
CNN14-BART &56.0 &16.0 &37.0 &17.1 &39.3 &11.7 &25.5\\
HTSAT-BART &\textbf{57.6} &\textbf{16.4} &\textbf{38.2} &\textbf{17.5} &\textbf{41.5} &\textbf{11.9} &\textbf{26.7}\\
Cacophony (Ours)& 50.8 &11.5&34.3&15.3  &34.2&10.6&22.4\\
\bottomrule
\end{tabular}
}
\end{table}
\vspace{-1cm}

\end{center}
\vspace{-0.2cm}
\subsection{Automated audio captioning}

\begin{table}[!t]
\caption{Representative examples of automatic captioning by Cacophony on Clotho dataset.}
\centering
\resizebox{0.85\columnwidth}{!}{
\begin{tabular}{c|c}
\toprule
\textbf{Ground Truth} &\textbf{Generated by Cacophony} \\
\hline
 \makecell{Plastic and other materials \\ rustle and crinkle continuously.}  &\makecell{A sound of a plastic \\ bag being crumpled.}\\
 \hline
\makecell{A quiet environment with a few \\insects making a sound and some \\birds chirping far away.}&\makecell{Crickets chirping \\in the background.} \\
\hline
\makecell{A door is being unlatched creaking \\ open and being fastened again.}& \makecell{A door opening\\ and closing.} \\
\bottomrule
\end{tabular}}
\label{tab:gen_cap}
\end{table}

\subsubsection{Task definition}
Automated audio captioning involves generating a free-form textual description for an audio signal, moving beyond predefined class labels or tags. 

\subsubsection{Experimental setup}
At inference, our model utilizes temperature sampling to generate captions at a temperature of 0.1.
The baselines apply beam search with a beam size of 3 for generating text.
We evaluate on the AudioCaps and Clotho datasets. 
For evaluation metrics, we use the Microsoft COCO Caption Evaluation package~\cite{chen2015microsoft}, which includes BLEU$_n$, ROUGE$_L$, METEOR, CIDEr, SPICE, and SPIDEr.

\subsubsection{Result}
We benchmark our captioning head against HTSAT-BART and CNN14-BART baselines proposed in~\cite{mei2023wavcaps} which are pretrained on WavCaps.
Both of these models have shown good performance in the automated audio captioning tasks on the Clotho and AudioCaps datasets.
We also include the synthetic audio captioner 
introduced in Sec.~\ref{sec:data}, HTSAT-BART-FT, which has been additionally fine-tuned on AudioCaps.

The results of this comparison are detailed in Table~\ref{tab:caption}.
In the AudioCaps dataset, our model exhibits competitive performance with that of the baseline models.
However, there is still a noticeable performance gap when compared to our ``teacher model'' HTSAT-BART-FT.
On the Clotho dataset, Cacophony's performance is notably lower than both CNN14-BART and HTSAT-BART. 
In Table~\ref{tab:gen_cap}, we include a few captiong examples of Cacophony that received low scores.
Looking at these examples, we identify two primary causes for our model's weak performance:
First, the textual output from Cacophony seems to show a different writing style from the ground-truths, lacking some 
detailed descriptions that are characteristic of Clotho's captions. 
It is expected that using a captioning model fine-tuned on Clotho to generate our pretraining data would enhance Cacophony's performance on this dataset, albeit potentially at the expense of performance on AudioCaps.
Second, Cacophony's audio encoder does not successfully detect all sound events. 
This is evidenced in the middle example, where the ``birds chirping far away" event is not correctly identified by our model. 


\begin{figure}[!t]
\centering
\includegraphics[width=0.85\columnwidth]{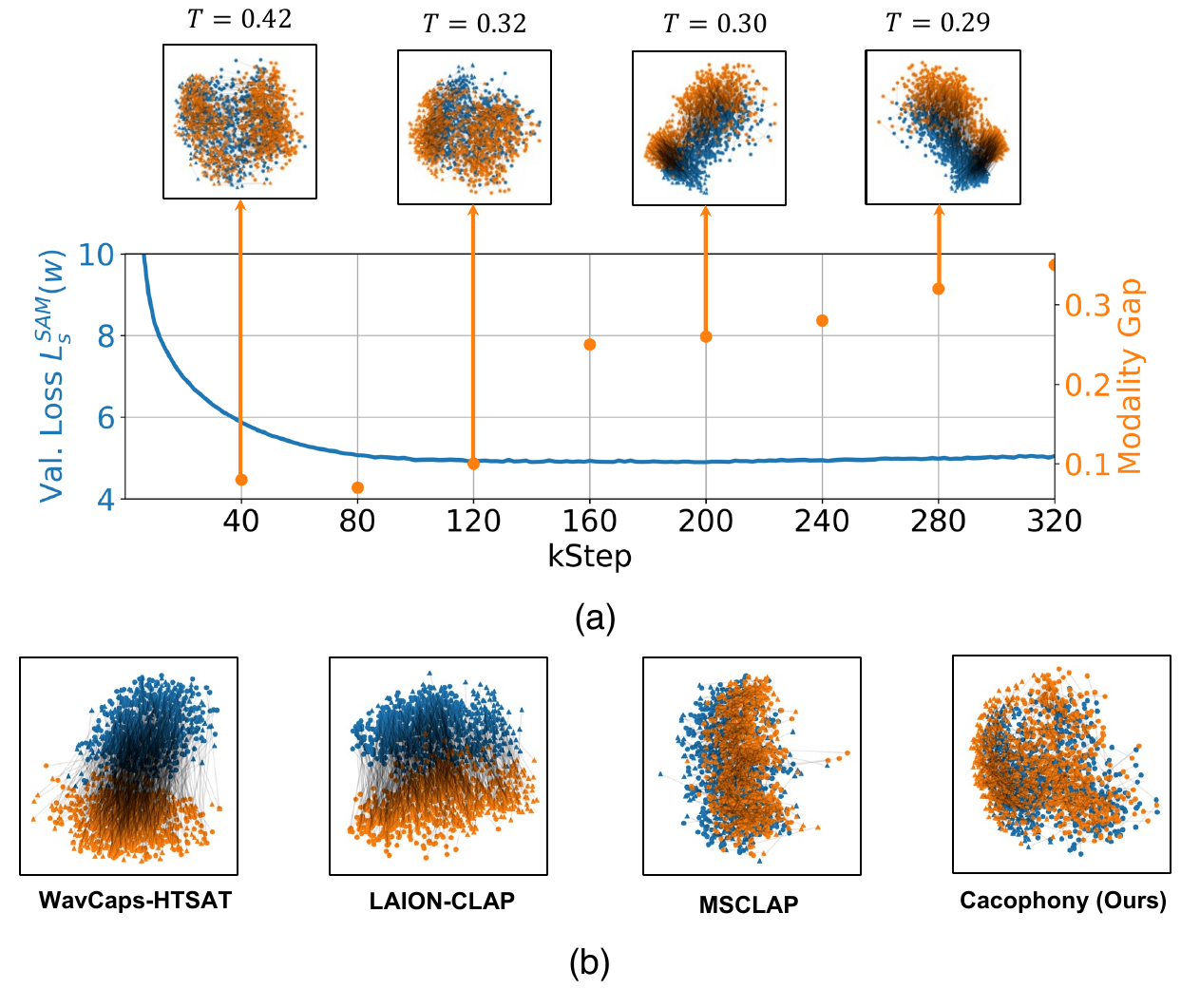}
    \caption{UMAP visualization of text and audio embeddings on the validation set. (a) Evolution of modality gap during our second-stage training, $T$ stands for the temperature parameter in Eq.~\ref{eq:con}. (b) Visualization of modality gap of existing models. }
    \label{fig:modgap}
\end{figure}

\subsection{Modality gap}
Although the contrastive training objective brings matched text-audio pairs closer than non-matched pairs, it does not necessarily bring audio and text into a joint multi-modal space globally.
Liang \textit{et al.}~\cite{liang2022mind} define the modality gap as the difference between the centroids of audio and text embeddings:
$\Delta_{gap}=\frac{1}{n}\sum_{i=1}^n \boldsymbol{x}_i -\frac{1}{n}\sum_{i=1}^n \boldsymbol{y}_i$,
where $\boldsymbol{x}_i$ and $\boldsymbol{y}_i$ denote the L2-normalized audio and text embeddings for the $i$-th sample in a given dataset.
The magnitude of this gap provides a quantitative measure of the global discrepancy between the audio and text embeddings.
This gap is shown to arise from the general inductive bias of deep neural architectures, random initialization and training objective~\cite{liang2022mind}.
Another factor may be the noisy correspondence of audio-text pairs, since correctly matched pairs guide the training, while mismatched pairs provide misleading supervision, where the pairs are incorrectly matched, as pointed out in Luong \textit{et al.}~\cite{luong2024revisiting}, where the correctly matched pairs guide the training, the noisy pairs incorrectly supervise the training.
As shown in contrastive image-text models, the existence of modality gap can impact its transferability to downstream tasks~\cite{liang2022mind}; we hypothesize that it may also be the case for contrastive audio-text models.
In our experiment, we track the evolution of the modality gap by plotting paired text and audio embeddings, as illustrated in Fig.~\ref{fig:modgap}a.
We observe that the modality gap begins to increase once the validation loss plateaus after 120k iterations.
When evaluating the text-audio retrieval task on the validation split, we find that checkpoints with smaller modality gap do not necessarily guarantee significantly better performance.
However, a model with a narrower gap may offer advantages in applications where embedding replacement across modalities is required, such as in text-to-audio generation~\cite{huang2023make,liu2023audioldm} or language-guided source separation~\cite{liu2023separate,dong2022clipsep}.
We also visualize the modality gaps of existing contrastive models, as shown in Fig.~\ref{fig:modgap}b. 
In models like LAION-CLAP and WavCaps-HTSAT, we notice that text and audio embeddings form distinct clusters. 

\vspace{-1cm}
\begin{center}
    \begin{table*}[!t]
  \caption{Ablations on text preprocessing, trained on openSFX, evaluated on test split of AudioCaps and Clotho.}
  \label{tab:ab_text}
  \centering
  \resizebox{0.9\textwidth}{!}{
  \begin{tabular}{c|ccc|ccc|ccc|ccc}
    \toprule
\multirow{3}{*}{\textbf{Recall ($\%$)}} & \multicolumn{6}{c|}{AudioCaps} & \multicolumn{6}{c}{Clotho} \\
    \cline{2-13}  & 
    \multicolumn{3}{c|}{Text-to-Audio} & \multicolumn{3}{c|}{Audio-to-Text} & \multicolumn{3}{c|}{Text-to-Audio} & \multicolumn{3}{c}{Audio-to-Text} \\
    \cline{2-13}
    & $\boldsymbol{R@1}$ & $\boldsymbol{R@5}$& $\boldsymbol{R@10}$  & $\boldsymbol{R@1}$ & $\boldsymbol{R@5}$& $\boldsymbol{R@10}$  & $\boldsymbol{R@1}$ & $\boldsymbol{R@5}$ &$\boldsymbol{R@10}$  &$\boldsymbol{R@1}$ & $\boldsymbol{R@5}$& $\boldsymbol{R@10}$  \\
    \hline
    None& 32.5&67.5&80.1&45.4&\textbf{76.5}&86.9&14.2&37.2&51.3&21.2&\textbf{51.9}&\textbf{66.2}\\
    LLM cleaning &\textbf{33.1}&\textbf{68.8}&\textbf{80.9}&\textbf{45.7}&76.1&86.0&17.2&41.2&55.2&\textbf{24.8}&51.2&65.8\\
    Recaptioning & 32.6&66.8&79.4&42.6&75.6&\textbf{88.1}&\textbf{17.7}&\textbf{43.0}&\textbf{56.9}&22.9&51.8&65.4\\
    \bottomrule
  \end{tabular}
  }
\end{table*}
\end{center}

\begin{center}
    \begin{table*}[!t]
  \caption{Ablations on the use of dataset for different stage training, $V_x$-$V_y$ represents that $V_x$ dataset is used for 1st stage MAE training and $V_y$ dataset is used for 2nd stage captioning contrastive training.}
  \label{tab:ab_data}
  \centering
  \resizebox{0.9\textwidth}{!}{
  \begin{tabular}{c|c|ccc|ccc|ccc|ccc}
    \toprule
\multirow{3}{*}{\textbf{Index}}&\multirow{3}{*}{\textbf{Recall ($\%$)}} & \multicolumn{6}{c|}{AudioCaps} & \multicolumn{6}{c}{Clotho} \\
    \cline{3-14}  & 
    &\multicolumn{3}{c|}{Text-to-Audio} & \multicolumn{3}{c|}{Audio-to-Text} & \multicolumn{3}{c|}{Text-to-Audio} & \multicolumn{3}{c}{Audio-to-Text} \\
    \cline{3-14}
    && $\boldsymbol{R@1}$ & $\boldsymbol{R@5}$& $\boldsymbol{R@10}$  & $\boldsymbol{R@1}$ & $\boldsymbol{R@5}$& $\boldsymbol{R@10}$  & $\boldsymbol{R@1}$ & $\boldsymbol{R@5}$ & $\boldsymbol{R@10}$ & $\boldsymbol{R@1}$ & $\boldsymbol{R@5}$& $\boldsymbol{R@10}$  \\
    \hline
    (a)&$V_1$ - $V_1$ &33.1&68.8&80.9&45.7&76.1&86.0&17.2&41.2&55.2&24.8&51.2&65.8\\
    (b)&$V_2$ - $V_1$ &32.4&66.9&79.2&42.7&76.0&85.8&15.7&41.0&55.1&21.5&49.5&64.0\\
    (c)&$V_3$ - $V_1$ &34.5&67.9&80.5&46.0&76.1&86.1&17.0&42.1&56.0&22.7&51.1&65.6\\
    (d)&$V_2$ - $V_2$ &31.6&64.2&77.7&44.4&78.3&87.7&18.6&42.7&56.8&25.0&52.2&66.3\\
    (e)&$V_3$ - $V_2$ &\textbf{40.5}&\textbf{75.6}&86.2&55.5&\textbf{84.4}&91.8&19.9&44.3&57.3&24.8&52.2&66.1\\
    (f)&$V_3$ - $V_3$ &\textbf{40.5}&75.2&\textbf{86.7}&\textbf{55.8}&83.1&\textbf{92.1}&\textbf{20.1}&\textbf{44.6}&\textbf{58.0}&\textbf{28.8}&\textbf{53.9}&\textbf{67.9}\\
    (g)&$\emptyset$ - $V_3$ &39.9&74.1&85.7&54.2&\textbf{84.3}&92.0&17.2&41.2&54.4&25.0&50.3&65.0\\

    \bottomrule
  \end{tabular}
}
\end{table*}
\end{center}

\begin{center}
\begin{table}[!t]
  \caption{Detailed information of training datasets in the dataset scale ablation study.}
  \label{tab:data_config}
  \centering
  \resizebox{\columnwidth}{!}{
  \begin{tabular}{c|c|c|c|c}
    \toprule
\textbf{Dataset} & Clean Labeled & Noisy-Labeled& Weakly/Un-Labeled & Duration\\
&&& & (kHour)\\
    \hline
    $V_1$ &AudioCaps, Clotho, OpenSFX&Freesound&-&4.2\\
    $V_2$&AudioCaps, Clotho, OpenSFX&Freesound&ACAV2M&7.8\\
    $V_3$&AudioCaps, Clotho, OpenSFX&Freesound&AudioSet, ACAV2M &13.2\\
    \bottomrule
  \end{tabular}
}
\end{table}

\end{center}
\vspace{-0.5cm}

\begin{center}
    \begin{table*}[!t]
  \caption{Ablations on different network configurations and initialization.}
  \label{tab:ab_net}
  \centering
  \resizebox{0.9\textwidth}{!}{
  \begin{tabular}{c|l|ccc|ccc|ccc|ccc}
    \toprule
\multirow{3}{*}{\textbf{Index}}&\multirow{3}{*}{\textbf{Recall ($\%$)}} & \multicolumn{6}{c|}{AudioCaps} & \multicolumn{6}{c}{Clotho} \\
    \cline{3-14}  & 
    &\multicolumn{3}{c|}{Text-to-Audio} & \multicolumn{3}{c|}{Audio-to-Text} & \multicolumn{3}{c|}{Text-to-Audio} & \multicolumn{3}{c}{Audio-to-Text} \\
    \cline{3-14}
    && $\boldsymbol{R@1}$ & $\boldsymbol{R@5}$& $\boldsymbol{R@10}$  & $\boldsymbol{R@1}$ & $\boldsymbol{R@5}$& $\boldsymbol{R@10}$  & $\boldsymbol{R@1}$ & $\boldsymbol{R@5}$ & $\boldsymbol{R@10}$ & $\boldsymbol{R@1}$ & $\boldsymbol{R@5}$& $\boldsymbol{R@10}$  \\
    \hline
    (a)&Cacophony &\textbf{40.5}&\textbf{75.2}&\textbf{86.7}&\textbf{55.8}&\textbf{83.1}&\textbf{92.1}&\textbf{20.1}&\textbf{44.6}&\textbf{58.0}&\textbf{28.8}&\textbf{53.9}&\textbf{67.9}\\
    (b)&\textit{- w/o} Captioning head & 38.9 & 73.2&85.4&54.5&82.0& 91.3&18.6& 42.8&56.6& 27.6& 52.2&64.5\\
    (c)& \textit{- w/o} SAM & 37.1 & 71.1&83.3&48.7&77.7&88.9& 15.5& 37.9&51.8& 19.6& 42.8&55.1\\
    (d)& \textit{- w/} Fixed Length &- &-&-&-&-&-& 19.1& 43.8&57.4& 27.8& 52.8&67.1\\
    \bottomrule
  \end{tabular}
  }
\end{table*}
\end{center}
\vspace{-1.2cm}
\section{Ablation studies}

In this section, we use audio-text retrieval for the ablation studies, as the task aligns with our pretraining contrastive objective.
In each ablation study, we use a 12-layer ViT-Base (ViT-B) Transformer with a hidden size of 512 and intermediate size of 1024 as our audio encoder in both training stages.
The other training hyper-parameters remain the same as those we described in Sec.~\ref{sec:train}.
At inference time, we choose to make use of the full sequence length.

\subsection{Text cleaning}
We first explore text pre-processing methods for the contrastive-captioning training stage, using only the Freesound, AudioCaps, and Clotho datasets.
To be specific, the compared text pre-processing methods include:
(1) implementing LLM cleaning as detailed in Sec.~\ref{sec:noisydata}; (2) conducting recaptioning based solely on audio, as described in Sec.~\ref{sec:weakdata}; (3) directly using raw text within a predefined maximal text length. 
For recaptioning the Freesound dataset (2), we use CNN-14-BART from WavCaps~\cite{mei2023wavcaps}, pretrained on WavCaps and then fine-tuned on Clotho, as its domain matches with Freesound.

The experimental result can be found in Table~\ref{tab:ab_text}. 
When compared to the raw text baseline, LLM cleaning generally improves performance for both text-to-audio and audio-to-text retrieval tasks on Clotho datasets, and achieves similar performance on AudioCaps dataset. 
Recaptioning yields mixed results; it improves performance on the Clotho dataset but slightly reduces accuracy on AudioCaps dataset.

For both text preprocessing methods, improvements are more consistent and significant in Clotho than in AudioCaps. 
We believe that Clotho evaluation is a more reliable indicator of the overall performance than AudioCaps evaluation, as the training dataset used in this ablation aligns closely with Clotho and is out-of-distribution for AudioCaps. 

\subsection{Dataset scale}
We perform an ablation study on the effect of training dataset size on model performance in the audio-text retrieval tasks (see Table~\ref{tab:ab_data}). 
To do so, we create three datasets 
designated as $V_1$, $V_2$ and $V_3$ with increasing sizes.
Detailed information of these datasets is provided in Table~\ref{tab:data_config}.

We first study the effect of dataset scale in the first-stage training.
In Table~\ref{tab:ab_data}), each of the three groups, (b-c), (d-e) and (f-g), uses a different dataset in the first-stage training but the same dataset in the second-stage training. 
In particular, Group (f-g) constitutes the extreme case: (f) uses the entire dataset in the first-stage training and (g) bypasses the first stage entirely.
Observations show a performance improvement across all tasks and datasets as the first-stage training data expands.
This evidences the positive impact of using larger datasets during the first-training stage on the model's overall effectiveness.
There are, however, two exceptions:
There is no noticeable improvement in retrieval performance when comparing (a) and (c), and there is a decline in all evaluation metrics when additional data from ACAV2M is incorporated when comparing (a) and (b). 
This lack of improvement in (b) and (c) compared to (a) may be due to the evaluation dataset being out-of-distribution of the limited training dataset $V_1$ used in the second stage.


We now analyze effects of data scale in the second-stage training, i.e., contrastive-captioning training.
In the comparison of experiments (c), (e), and (f), the audio encoder is initialized with weights from training on the $V_3$ dataset.
During the second stage, these groups are trained on $V_1$, $V_2$, and $V_3$, respectively. 
When evaluating on AudioCaps, there is a significant increase from (c) to (e) in recall when integrating ACAV2M ($V_2$) into the second-stage training, but this improvement appears to plateau from (e) to (f) upon the inclusion of AudioSet ($V_3$).
However, on Clotho, the retrieval performance consistently improves with increasing dataset scale from (c) to (e) and (f), without observation of a plateau, showing the effectiveness of increasing data scale.

In experiments (a), (d), and (f), we increase the dataset scale simultaneously for both training stages and observe a gradual improvement on retrieval performance on the Clotho test dataset. 
However, a significant improvement in retrieval performance on AudioCaps is observed only when integrating AudioSet (group (f)) into both training stages. 

We find scaling up the datasets in both stages tends to lead to improved model performance.
The MAE training step contributes to the better retrieval performance for the audio-text model. 
Therefore, collecting more unlabeled audio or using larger versions of the ACAV dataset for audio encoder MAE training could potentially yield further gains in downstream performance.


\begin{center}
\begin{table}[!t]
\caption{ 
Classification accuracy ($\%$) on long-sequence classification from HEAR benchmark.
``Fixed length" refers to randomly sampling a 10-second patch to match the audio duration in training.
``Full length" refers to using the full audio.
}
\label{tab:len}
\centering
\resizebox{0.95\columnwidth}{!}{
\begin{tabular}{c|c|c|c}
\toprule
Dataset & Duration (sec) & Fixed Length &Full Length\\
\hline
GTZAN-Genre &30& 83.7& 85.4
\\
GTZAN-Music Speech&30&97.7&98.5\\
VoxLingua 107&18.6 &25.0 &26.5\\
\bottomrule
\end{tabular}
}
\end{table}
\end{center}
\vspace{-1cm}
\subsection{Architectural design}
Regarding the neural network architecture, we examine the effect of incorporating a captioning head, SAM optimization, and fixed length processing window on audio-text retrieval tasks (see Table~\ref{tab:ab_net}).
For these experiments, we use the full-size dataset $V_3$ in both training stages, i.e., setting (f) in Table~\ref{tab:ab_data}.
We further demonstrate the adaptability of our audio encoder in handling varying audio lengths.

\subsubsection{Captioning Objective.} By comparing experiments (a) and (b) in Table~\ref{tab:ab_net}, we observe a marginal improvement across both test sets when the captioning objective is included.
We hypothesize that the captioning objective facilitates the learning of more fine-grained audio representations, which aligns with the findings of BLIP~\cite{li2022blip}, CoCa~\cite{yu2022coca} and CapCa~\cite{tschannen2023image}.

\subsubsection{Use of SAM.} 
By comparing (a) and (c), we see that SAM significantly improves the accuracy on the audio-text retrieval task for both datasets.
However, the benefits of SAM come at the expense of two sequential gradient computations during each training step; as mentioned earlier, there are methods for reducing SAM computational overhead that could be integrated in future work.  


\subsubsection{Flexibility to Length.}
Lastly, we explore the capability of our model to process longer audio during inference.
We start by evaluating the audio-text retrieval task on the Clotho dataset, where audio sample duration ranges from 15 to 30 seconds. 
We compare the performance between (a), which uses the entire length of the audio, and (e), which randomly samples 10-second spectrogram MAE patches from the full audio to match the training duration.
We find that using longer sequence lengths leads to an improvement in retrieval recall, even though our training only uses 10-second time-frequency patches.
Then, we extend our investigation to include long sequence classification tasks from the HEAR benchmark, specifically focusing on the GTZAN~\cite{tzanetakis2002musical} and VoxLingua~\cite{valk2021voxlingua107} tasks, as shown in Table~\ref{tab:len}. 
Our findings reveal that leveraging the full length of the audio samples consistently yields higher classification accuracy compared to using fixed length patches, which is consistent with findings in  FLIP~\cite{li2023scaling}.

\IEEEpubidadjcol
\vspace{-0.1cm}
\section{Conclusions}


In this paper, we presented dataset and model improvements for contrastive audio-language models.
By automatically captioning unlabeled audio data and refining noisy existing captions with language models, we curated a dataset of 3.9 million text-audio pairs. 
Our modeling approach consists of a two-stage training strategy: The first stage involves training an audio encoder with an MAE objective, while the second stage combines contrastive learning with an auxiliary captioning task. 
Our final model, Cacophony, achieves state-of-the-art performance on audio-text retrieval tasks and competitive performance on other downstream tasks.

We conducted a series of ablation studies to evaluate the effects of our dataset curation and modeling strategies. 
Notably, we observed positive effects of our modeling techniques and dataset scaling in both training phases. The improvement seen from scaling MAE pretraining data is particularly relevant, as MAE training leverages audio-only data and can easily scale further in future work.

While our work demonstrates significant advancements in audio-text modeling, it is important to acknowledge several limitations. 
We observe weaker performance in fine-grained classification and captioning tasks, suggesting the need for more detailed, higher-quality captions in our training data. 
In particular, the reliance on a single-modality audio captioner for synthesizing a large portion of our data may limit the diversity and richness of our dataset. 
Integrating multimodal captioning synthesis techniques, such as those involving the visual modality, could potentially address this limitation.
Furthermore, our model's relatively weak performance on speech-related tasks indicates a need for more comprehensive speech data integration. 
Finally, the persistent modality gap between audio and text embeddings, for our proposed model and other audio-text models, remains a challenge, potentially limiting performance in tasks requiring tight cross-modal alignment.






\bibliographystyle{IEEEtran}
\bibliography{references}

 


\end{document}